\documentclass[aps,prb,amsmath,amssymb,footinbib,showpacs,superscriptaddress]{revtex4-1}
\usepackage{newtxtext,newtxmath}
\usepackage[letterpaper,margin=1in]{geometry}
\usepackage{amsmath}
\usepackage{dsfont}
\usepackage{graphicx}
\usepackage{dcolumn}
\usepackage{bm}
\usepackage{textcomp}
\usepackage{color}
\usepackage[normalem]{ulem}

\begin{document}

\title{Temperature Dependent Angular Dispersions of Surface Acoustic Waves on GaAs}
\author{Mats Powlowski} 
\affiliation{Institute for Quantum Computing, University of Waterloo, 200 University Avenue West Waterloo, ON, N2L 3G1, Canada}
\affiliation{Department of Electrical and Computer Engineering, University of Waterloo, 200 University Avenue West, Waterloo, ON, N2L 3G1, Canada}
\author{Francois Sfigakis}
\affiliation{Institute for Quantum Computing, University of Waterloo, 200 University Avenue West Waterloo, ON, N2L 3G1, Canada}
\author{Na Young Kim}
\affiliation{Institute for Quantum Computing, University of Waterloo, 200 University Avenue West Waterloo, ON, N2L 3G1, Canada}
\affiliation{Department of Electrical and Computer Engineering, University of Waterloo, 200 University Avenue West, Waterloo, ON, N2L 3G1, Canada}
\affiliation{Department of Physics and Astronomy and Department of Chemistry, University of Waterloo, 200 University Avenue West, Waterloo, ON, N2L 3G1, Canada}
\affiliation{Perimeter Institute, 31 Caroline Street North, Waterloo, ON, N2L 2Y5, Canada}
\email{Corresponding author: nayoung.kim@uwaterloo.ca} 

\begin{abstract}
We measure the phase velocities of surface acoustic waves (SAWs) propagating at different crystal orientations on (001)-cut GaAs substrates and their temperature dependance. We design and fabricate sets of interdigital transducers (IDTs) to induce 4 $\mu$m SAWs via the inverse piezoelectric (PZE) effect between the PZE [110] direction (set as $\theta$ = 0\textdegree) and the non-PZE [100] direction ($\theta$ = 45\textdegree) on GaAs. We also prepare ZnO-film sputtered GaAs substrates in order to launch SAWs efficiently by IDTs even in the non-PZE direction. We quantify acoustic velocities between 1.4 K and 300 K from the resonant frequencies in the $S_{11}$ parameter  using a network analyzer. We observe parabolic velocity-temperature trends at all $\theta$-values both on GaAs and ZnO/GaAs substrates. Below 200 K, in ZnO/GaAs substrates slower SAW modes appear around the [110] direction, which are unseen at room temperature.
\end{abstract}

\maketitle


How to store, transduce, and process information is at the heart of communications and computation. A popular means for conveying information is electromagnetic waves (photons) featuring insignificant loss due to little interaction with environment. Various photonic structures serve as optical interfaces between photons and information hosts in solids. Another information transduction method exploits mechanical vibrations of lattices in solids, so-called phonons\cite{Kittel}. Besides natural crystal vibrations, acoustic phonons can be generated  from the crystal's elastic responses to external forces in the form of stress and strain.  Short laser pulses cause materials' thermal expansion and contraction, giving rise to acoustic waves \cite{White1963}. Alternatively, the inverse piezoelectric  (PZE) process offers a convenient method to generate acoustic waves by modulating electrical signals through devices such as interdigital transducers (IDTs)\cite{White1965}.

Among several types of acoustic longitudinal and transverse waves in solids, Rayleigh waves exist on the interface between free space and the surface. They are known as surface acoustic waves (SAWs). They propagate distinctively either (1) without angular dispersion in an isotropic PZE material, such as $c$-axis oriented hexagonal ZnO\cite{Dybwad1971,Deger1998} or (2) dispersively in anisotropic PZE materials, such as zinc-blende GaAs\cite{Flannery1999, Deacon1972}.  Angular dispersions are related to diffraction of SAWs on surface, for example, a downward parabolic dispersion on GaAs leads to autocollimating behavior of SAWs\cite{Kim1994}.   
Depending on propagation direction in GaAs, there are two different SAW modes\cite{Flannery1999, Deacon1972}: (1) pure Rayleigh SAWs and (2) pseudo or leaky SAWs denoted as pSAWs. pSAWs are partially localized near the surface, but they lose energy into the bulk by transverse shear waves\cite{Hunt1986}. While faster pSAWs appear around the PZE [110] direction on (100)-cut GaAs, slower SAWs  propagate around the non-PZE [100] direction at room temperature\cite{Flannery1999, Deacon1972}. 

Periodically oscillating electric fields are generated on PZE materials when RF or microwave signals are applied to IDTs. The interlocking periodic metal electrodes of IDTs directly convert electric energy to acoustic elastic waves with acoustic wavelength equal to the electrode period\cite{White1965}. Since the wave phase velocity $v$ is material-characteristic, the desired SAW wavelength $\lambda_\text{SAW}$ is determined by the simple relation $v = f \lambda_\text{SAW}$, with driving frequency $f$. Consequently, the period of the metal electrodes and the spacing of the fingers can be designed. However, the [100] direction on (001)-GaAs is non-PZE, which means that SAWs cannot be produced by IDTs. 
This intrinsic material limitation is overcome by depositing a thin film of strong PZE material, such as ZnO, where
the ratio of the PZE film thickness to $\lambda_\text{SAW}$ and the substrate material influence the overall velocity values of SAWs in this hybrid system\cite{Du2008}.  

Straightforward IDT design makes SAWs useful in many applications. Classically, SAWs have been used for acoustic cavities \cite{Schuetz2015}, RF filters\cite{Yoshino2000}, delay lines\cite{Adkins1971}, and pressure\cite{Jiang2005}, chemical\cite{Kepley1992} and biosensors\cite{Fu2017}. In the field of quantum technologies, SAWs have been used to store and transport electrons\cite{Cunningham1999}, spins\cite{Sogawa2001}, excitons\cite{Alsina2004, Rudolph2007} and exciton-polaritons\cite{Cerda-Mendez2011}, to modulate quantum-dot energy levels\cite{Villa2017} and to couple to superconducting qubits\cite{Manenti2017}. In addition, a universal quantum data bus based on SAWs\cite{Schuetz2015} is proposed for quantum information processing, and fundamental description of  quantum acoustics via SAWs has been examined\cite{Aref2016}. Two-dimensional SAW lattices for electrons in semiconductors are suggested for studying quantum many-body physics\cite{Byrnes2007, Schuetz2017}. Often these quantum systems operate at cryogenic temperatures. Despite the growing use of SAWs in GaAs and hybrid ZnO/GaAs in classical and quantum applications, a systematic essential study of SAW velocity dispersions and their temperature dependence is so far lacking.  

Here we report angular dispersions and their temperature dependence of SAWs on bare (001)-cut GaAs and ZnO-covered (001)-cut GaAs substrates. We choose ZnO because its SAW velocity ($\sim$ 2700 m/s) is close to the SAW velocity in GaAs ($\sim$ 2867 m/s in the PZE [110] direction).  A 316 nm-thick ZnO layer was sputtered using an RF magnetron sputtering system. The chamber pressure was 10 mTorr, maintained by a 27 sccm flow of a 60:40 Ar:O2 mixture. The substrate was kept at 100\textdegree C and rotated at 60 rpm, followed by annealing in O$_2$ at 300\textdegree C for 30 minutes. X-ray diffraction data show a dominant (002) ZnO peak, indicating that the desired $c$-axis orientation was grown on GaAs \cite{Hickernell1976}. 

\begin{figure}[!htbp]                          
\includegraphics[width=3.2in]{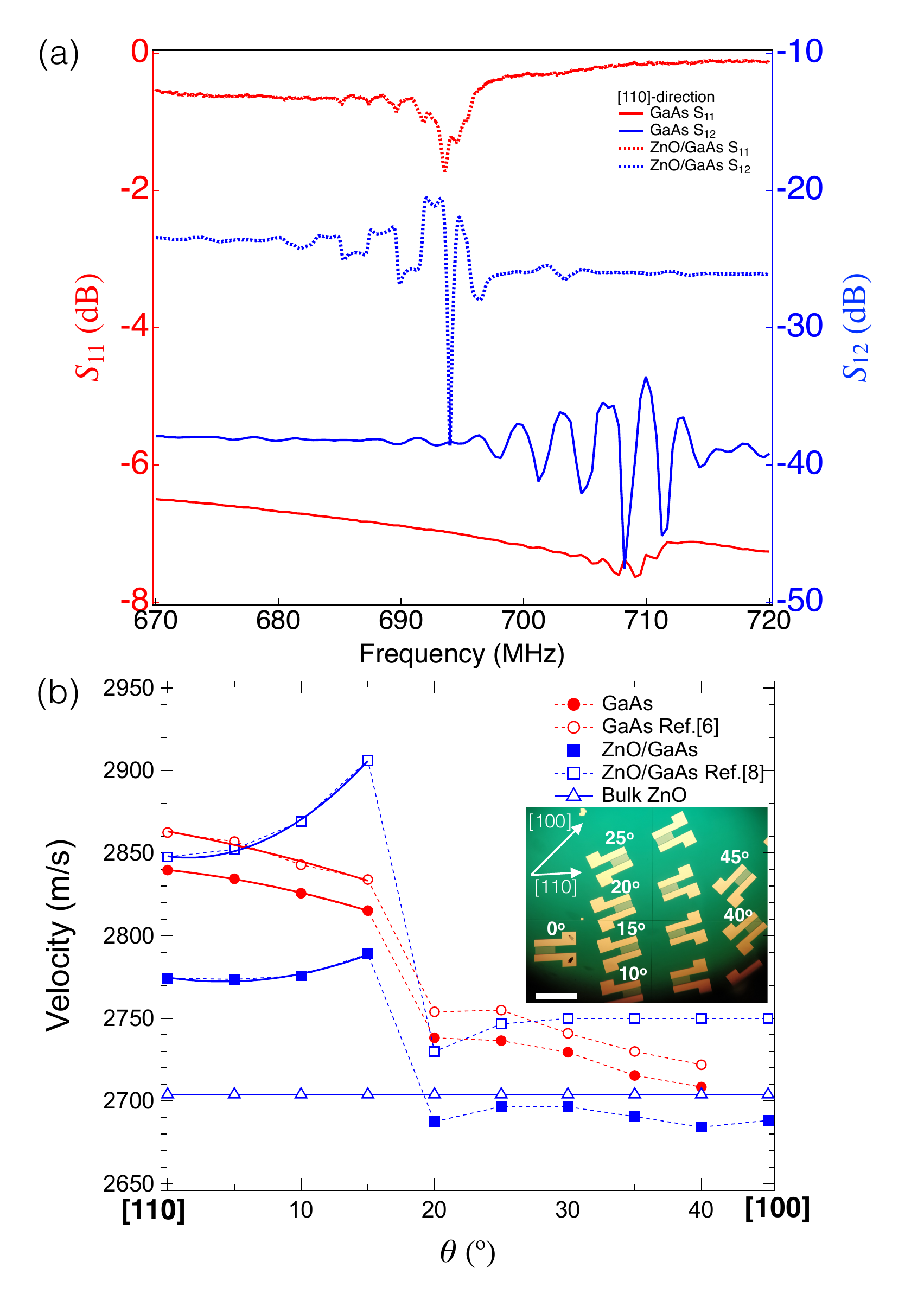}
\caption{(a) Room temperature $S_{11}$ and  $S_{12}$ data for propagation in [110] direction on GaAs and ZnO/GaAs samples. (b) Comparison of our experimental results at room-temperature to the reported values from literature for GaAs\cite{Flannery1999}, bulk ZnO and ZnO thin film on GaAs\cite{Kim1994}, as a function of angle $\theta$ from the GaAs [110] direction. Dashed lines are guides for the eye while solid lines are parabolic fits to the data. Our experimental results are taken from $S_{12}$-parameter measurements. Inset: A photo of the fabricated interdigital transducers (IDTs) on the ZnO/GaAs sample. Adjacent IDT pairs are separated by 5\textdegree$\>$. A scale bar is 600 $\mu$m-long.}
\label{fig:fig1}
\end{figure}

For the angle dependence, we fabricate sets of IDTs in 5\textdegree$\>$intervals from the PZE [110] to the non-PZE [100] direction shown in the inset of Fig. \ref{fig:fig1}(b). We set the GaAs [110] direction as the IDT angle $\theta$ = 0\textdegree$\>$, consequently that of the [100] direction as $\theta$ = 45\textdegree. Based on the zinc blende crystal symmetry and in agreement with previous measurements \cite{Flannery1999, Hunt1986}, we limit the angle range to be  $\theta \in \{0$\textdegree, 45\textdegree$\}$ since the velocity distribution between the [100] and [010] axes is symmetric with respect to the [110] direction. Our $\lambda_\text{SAW}$ = 4 $\mu$m IDTs ($\sim$ 700 MHz) are single-fingered with a metallization ratio 0.5, leading to a finger width of 1 $\mu$m and a finger-to-finger spacing of 1 $\mu$m, which can be readily achieved by photolithography and electron-beam lithography. Each IDT has 70 pairs of fingers and its aperture is 100 $\mu$m. Two IDTs forming a pair are separated by 500 $\mu$m. We deposit 20 nm/20 nm of Ti/Au for fingers and 20 nm/80 nm of Ti/Au for the busbars and wirebonding pads. 

We perform standard scattering $S$-parameter measurements using a vector network analyzer. 
The resonant frequencies $f$ appear at the return loss minima or transmission maxima frequencies identified from the $S_{11}$ or $S_{12}$ spectra, whose representative plots are presented in Fig. \ref{fig:fig1}(a), where the aforementioned formula is applied to quantify the velocities $v$. We note that the $S_{12}$ background signals at non-resonant frequencies are high, -25 dB for ZnO/GaAs and -40 dB for GaAs. We attribute those to come from our imperfect transmissions such as impedance of measurement coax cables and other electrical coupling paths in materials. Indeed, even in the same measurement setup, the$S_{12}$ background signal values of ZnO/GaAs are worse than those of GaAs. We think that the ZnO film quality from our sputtering process may not be as cleans as that of the GaAs wafers.  In addition, we perform the inverse Fourier transform of the $S_{12}$ spectra, from which we confirmed that our mechanical coupling associated with the SAW frequency is 10-100 times stronger than any spurious low frequency coupling. The ripples seen in the $S_{11}$ spectra may be caused by double transit, where a portion of the SAW energy is reflected by the output IDT. The 2 MHz ripple frequency may sit within the frequency range [1.34, 2.78] MHz due to the round-trip from the shortest distance between the first finger pair to the longest distance between the last finger pair. Thus, we think the majority of the SAW energy will come from closer to the centre of the IDTs.
Figure \ref{fig:fig1}(b) collects the angle-dependent velocity values from our room temperature (RT) measurements with both GaAs (filled red circle) and a 0.079$\lambda_\text{SAW}$ (316 nm) ZnO film on GaAs (filled blue square) from the $S_{12}$ measurements. We note that the resonant frequency $f$ values from the $S_{12}$ parameters differ by 1-2 MHz from those of the $S_{11}$ parameters, within a maximum of 0.29\% error, which is negligible. To extract the velocities, we apply the simple relation $v = f \lambda_\text{SAW}$ with the resonant frequencies from the $S$-parameter spectra without introducing any parameters such as acoustic reflection or temperature-dependent elastic response, which are not easily quantified. 

Our RT results (filled circle and filled square in Fig. \ref{fig:fig1}(b)) are compared to previous measurements made using laser pulses on GaAs\cite{Flannery1999} (open circle) and line-focus-beam scanning acoustic microscopy on the surface of the 0.13$\lambda_\text{SAW}$-thick ZnO ($\lambda_\text{SAW}$=1.6 $\mu$m)/GaAs substrate \cite{Kim1994}(open square). Dispersionless bulk ZnO data, found from COMSOL simulations, is shown in open triangles. Apart from the absolute values of our experimental velocities, differing by at most 22 m/s (0.8\%) for GaAs and 116.8 m/s (4.02\%) for ZnO/GaAs, the overall trends of the SAW angular dispersions in both substrates coincide with literature data. We attribute the bigger discrepancy in ZnO/GaAs between our result and the literature data to the different ZnO film thickness. Our ZnO layer is as thick as 0.079$\lambda_\text{SAW}$ (316 nm), thinner than the 0.13$\lambda_\text{SAW}$ ZnO layer (1.6 $\mu$m for $\lambda_\text{SAW}$ = 12 $\mu$m) used by Kim $\emph{et al.}$\cite{Kim1994}. 

We consider that the difference in measured velocity on GaAs in Fig. \ref{fig:fig1}(b) is due to the presence of the IDTs on the material surface. The PZE coupling in GaAs is typically 0.07\%\cite{Kim19942}, meaning that the velocity shift due to shorting of the surface is only 1 m/s. The metal fingers are 0.01$\lambda$ thick, the same as in Deacon and Heighway\cite{Deacon1972}; their results closely match ours (within 3 MHz or 0.45\% in most cases) but use Al electrodes. The difference in material densities does not make the masses equal, suggesting that mass loading is one of but not the only contributing factor. The sound velocity in Au is less than half that in GaAs, compared to Al and Ti, in which the sound velocity is slightly larger than GaAs; this has been used to show that Au electrodes decrease the measured velocity\cite{Takagaki2002}. Additional work would be necessary to determine how much each of these factors contributes to the lower measured velocity.

Two distinctive regions are separated in Fig.\ref{fig:fig1} (b) between $\theta$ = 15\textdegree$\>$and 20\textdegree$\>$ with both the bare GaAs and the ZnO/GaAs samples. In the region of $ 0 < \theta < 20$\textdegree,  the signals are assigned as pSAWs, which propagate faster than the SAWs assigned to  larger angles \cite{Hunt1986, Flannery1999, Deacon1972}. This behaviour is replicated in the ZnO/GaAs device, suggesting that the ZnO velocity depends heavily on the underlying GaAs substrate. This is to be expected, as a SAW should penetrate about 1$\lambda_\text{SAW}$, which corresponds to 4 $\mu$m, while the ZnO layer is only 0.079$\lambda_\text{SAW}$. In addition, both our and the literature ZnO/GaAs devices exhibit a strong concave-upward parabolic behavior except with different fitting coefficients, which is opposite to a mild concave-downward behavior in the GaAs device. Thus, the beam diffracts more in the ZnO/GaAs system, unlike in GaAs, where it is autocollimating\cite{Kim1994}. pSAWs suffer from the energy loss to the bulk, which reduces driving power efficiency. The $\theta$-dependent parabolic fit results are summarized in Table \ref{tab:table1}.

\begin{figure}
\includegraphics[width=3.25in]{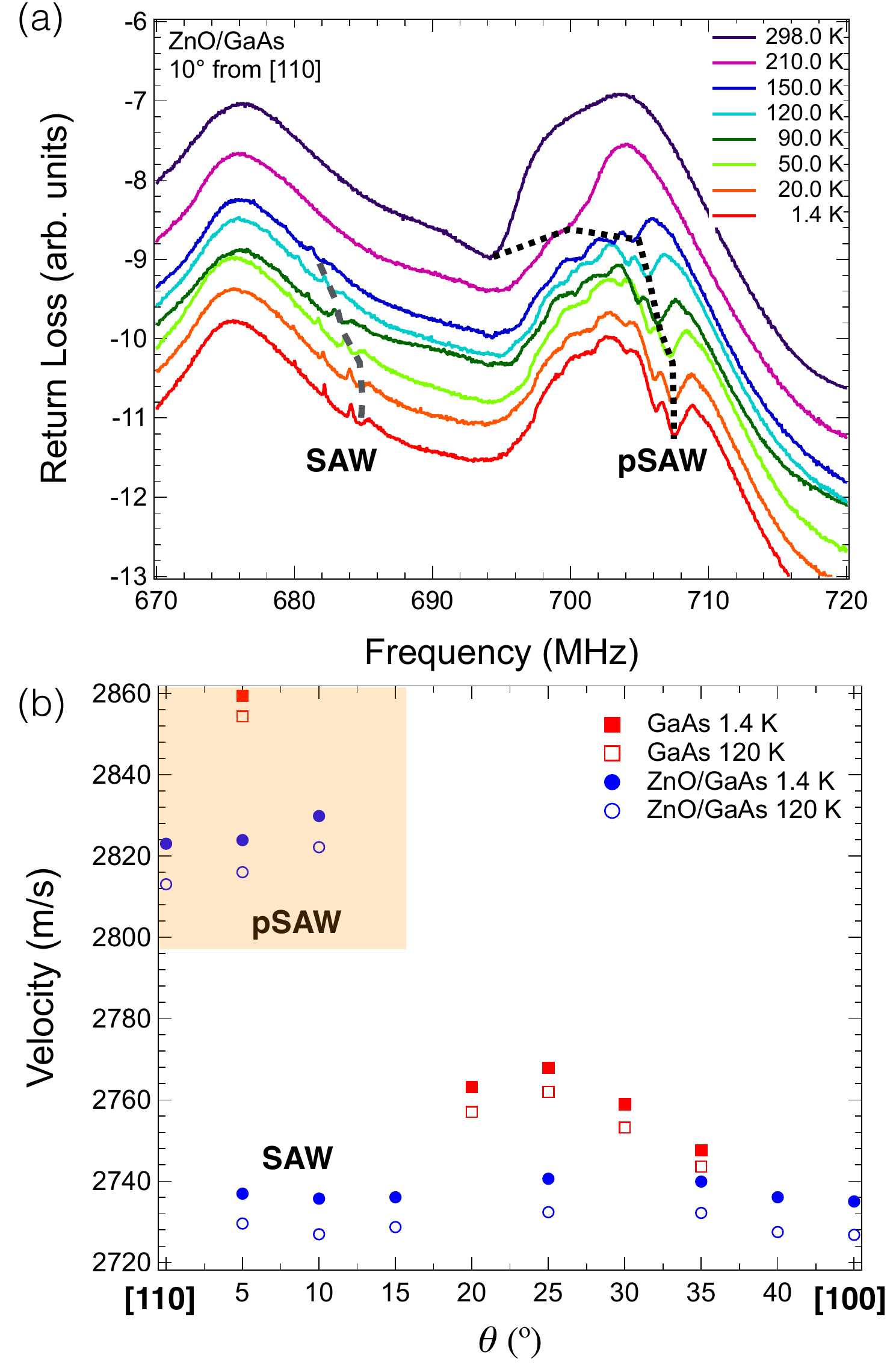}
\caption{(a) Return loss spectra of an IDT on the ZnO/GaAs device at $\theta$ = 10\textdegree$\>$, measured from 1.4 K to 150 K. The change in resonance around 680 MHz and 705 MHz is indicated by dashed lines, corresponding to the SAW and the pSAW modes in GaAs, respectively. Arbitrary vertical offsets are applied to each spectra for clarity.  Cables have not been calibrated out. (b) The velocities seen for bare GaAs and ZnO/GaAs at 1.4 K and 120 K.}
\label{fig:fig2}
\end{figure}

\begin{figure}[htbp]
\includegraphics[width=3.2in]{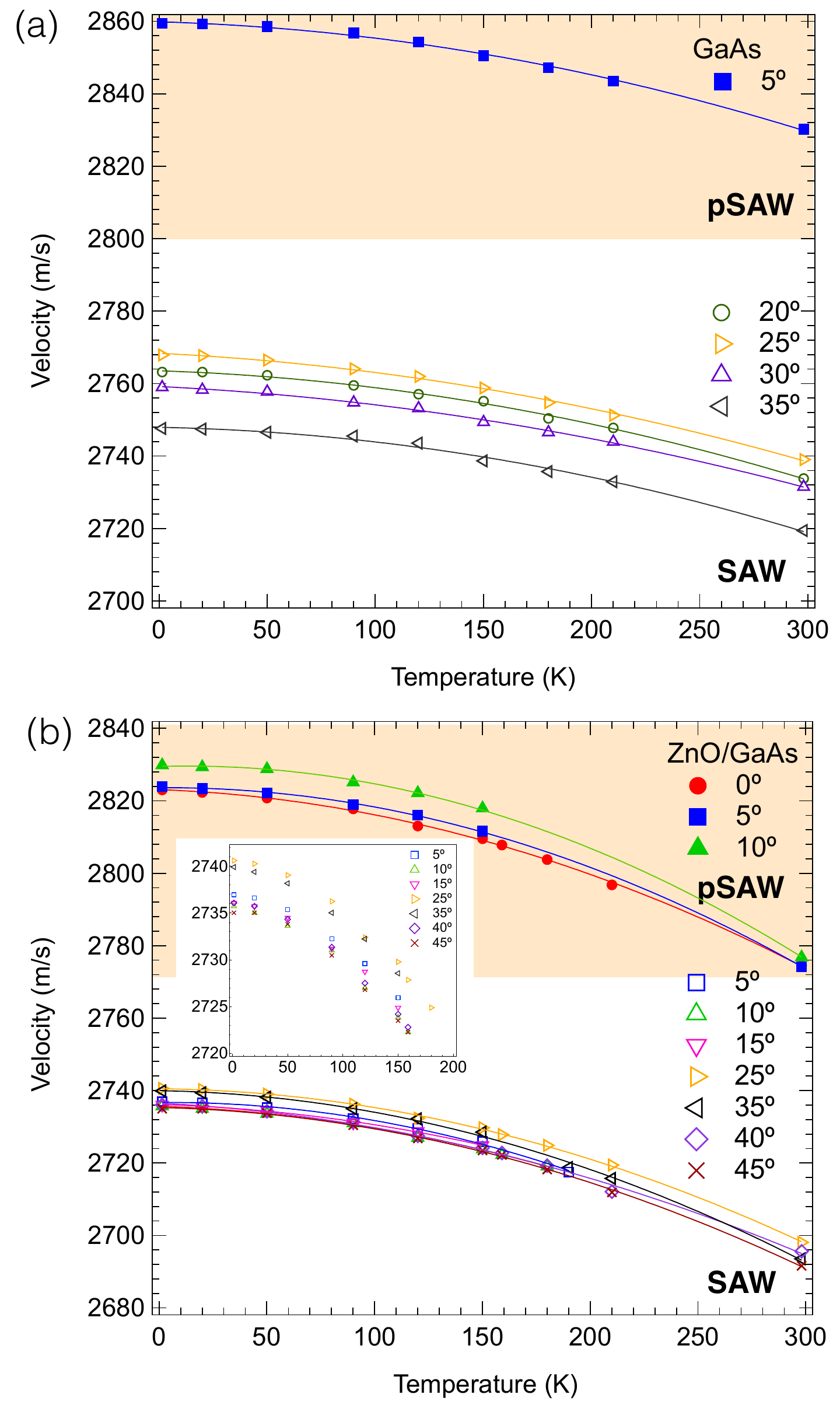}
\caption{Temperature dependence of the SAW and pSAW velocities on (a) GaAs and (b) ZnO/GaAs devices. Lines are parabolic fits to data. Maximum change from room temperature (RT) to 1.4 K is (a) 1.08\% and (b) 1.79\% of RT velocity for both pSAW and SAW groups. Inset in (b): Magnification of SAW data from 1.4 K to 200 K.}
\label{fig:fig3}
\end{figure}

For temperature-dependent measurements, we place our devices in a liquid-helium cryostat with six microwave coax cables, whose base temperature $T$ is 1.4 K and we take measurements after complete thermalization at each $T$. In order to maximize the number of devices per cool-down, we measure the $S_{11}$-parameters, which requires only half the number of coax cables per device. We collect the return-loss spectra from all devices in GaAs and ZnO/GaAs substrates through five cool-downs in total. We note that some signals for several IDTs are too small to be visible, so we are not able to obtain data from all devices. These may be possibly due to impedance mismatch at low temperatures. 

Figure \ref{fig:fig2}(b) presents representative low temperature angular dispersions of GaAs (red) and ZnO/GaAs (blue) substrates at 1.4 K (closed symbols) and 120 K (open symbols). The angular dispersion behaviors look similar at low temperatures to the RT dispersions; namely, there are regions of faster pSAWs and slower SAWs separated around 20\textdegree. In addition, we notice in ZnO/GaAs devices that slower SAW modes appear as well as pSAWs in the smaller $\theta$ region. The $S_{11}$ spectra of an IDT on ZnO/GaAs at $\theta$ = 10\textdegree$\>$ are shown as an example in Fig. \ref{fig:fig2}(a), where two unique resonant signals are seen. Based on the velocity values, we associate the lower and higher frequency signals with SAW and pSAW components, respectively. The dashed lines in Fig. \ref{fig:fig2}(a) show the points taken as resonance frequencies for the SAW and pSAW components, which are plotted to scale in Fig. \ref{fig:fig3}(b). In particular, at $T$ = 120 K, the two resonant $f$ values at $\theta$ = 10\textdegree$\>$ are located at 681.962 MHz and 705.548 MHz, corresponding to $v$ of 2727.85 m/s and 2822.19 m/s, respectively.  Since the velocity of the SAW modes near the [110] direction is rather constant, these modes in ZnO/GaAs may not diffract like pSAWs. These SAW modes do not appear above 200 K in bulk ZnO crystals, possibly due to an decreased resistivity in ZnO, which may damp SAW modes \cite{Magnusson2015}. However, not all SAW modes of the ZnO/GaAs devices disappear at low temperatures, consequently this feature provides a more broadly applicable system. For example, dispersionless SAWs in ZnO/GaAs at low temperatures may make the SAW propagation beam straightened. As both GaAs and ZnO/GaAs samples cool down, the velocity values of pSAWs and SAWs increase. 

The $T$-dependent velocities of SAWs and pSAWs are plotted for GaAs (Fig. \ref{fig:fig3}(a)) and ZnO/GaAs (Fig. \ref{fig:fig3}(b)) at specific angles. All data are fitted well with parabolic relations, and the exemplary fitting results are included in Table \ref{tab:table1}. Since the difference in velocity values is consistent at different $\theta$ values (Fig. \ref{fig:fig2}(b)), we are able to interpolate $v$ values at any temperatures by applying appropriate pSAW or SAW parabolic relations to the RT velocity data. Knowledge of the parabolic $T-v$ trends facilitates  design of SAW and pSAW devices operating at any temperatures and any orientations, for a ZnO thickness ratio close to 0.08$\lambda$.

\begin{table}[t]
\caption{\label{tab:table1}Fitting Parameters for parabolic relations  $v = a\theta^2 + b\theta + c$ or $v = aT^2 + bT + c$ to $\theta$ and $T$ dependent data. The $\theta$-data are taken between 0\textdegree and 20\textdegree in Fig. \ref{fig:fig1}(b). For the $T-v$ data, pSAW and SAW data are selected at $\theta$ = 5\textdegree and 25\textdegree  respectively from Fig. \ref{fig:fig3}.}
\begin{tabular}{lllll} 
\hline\hline
Data&
\multicolumn{1}{c}{$a$}&
\multicolumn{1}{c}{$b$}&
\multicolumn{1}{c}{$c$}\\
\hline
GaAs ($\theta$) \cite{Flannery1999}& -0.035 & -1.46 & 2863.2\\
Exp. GaAs ($\theta$) & -0.0525 & -0.858 & 2839.8\\
ZnO/GaAs ($\theta$) \cite{Kim1994} & 0.324 & -1.01 & 2848.1\\
Exp. ZnO/GaAs ($\theta$) & 0.138 & -1.15 & 2774.8\\
\hline
pSAW GaAs ($T$) & -0.00028 & -0.017 & 2859.9\\
SAW GaAs ($T$) & -0.00023 & -0.030 & 2759.2\\
pSAW ZnO/GaAs ($T$) & -0.00057 & 0.0035 & 2823.6\\
SAW ZnO/GaAs ($T$) & -0.00045 & -0.0066 & 2740.5\\
\hline\hline
\end{tabular}
\end{table}

We believe that our systematic quantification of SAW phase velocities will open up interesting applications with SAWs at arbitrary directions in GaAs and hybrid ZnO/GaAs at low temperatures. In particular, the dispersionless SAW modes in ZnO/GaAs below 200 K would be of use for propagation without much diffraction. We think that our work provides fundamental material parameters to develop phenomenological models based on coupling of modes \cite{Brown1989} or impulse response\cite{Hartmann1988} for SAW devices operating at different propagation directions with respect to the crystal orientation.

\begin{acknowledgments}
We thank N. N. Fitzpatrick for valuable advice on nanofabrication in Quantum NanoFab, K. Choi and S. Motlakunta for sample holder machining and C. Deimert and Z. Wasilewski for providing GaAs substrates. We thank the Canada First Research Excellence Fund for their equipment and facilities to be used for this study. We also thank CMC Microsystems provided financial assistance for devicefabrication. This work is supported by NSERC Discovery Grant RGPIN-05034-2017, the Ontario Ministry of Research \& Innovation through Early Research Awards, the Canada Foundation for Innovation, Industry Canada and Mike \& Ophelia Lazaridis. M. P. is supported by Richard and Elizabeth Madter graduate award and NSERC CGS-M scholarship. 
\end{acknowledgments}

\end{document}